\documentclass{aa}
\usepackage{graphics}
\usepackage{times}

\newcommand{\cf}{{\em cf.~}}

\newcommand{\eg}{{e.g.~}}

\newcommand{\Teff}{$T_{\rm eff}$~}
\newcommand{\logg}{$\log g$~}
\newcommand{\loggf}{$\log gf$~}

\newcommand{\tauc}{$\tau_{\rm c}$~}

\newcommand{\Mag}{$^{24}$Mg}
\newcommand{\Sings}{3s~$\!^1$S}
\newcommand{\Singp}{3p~$\!^1$P$^{\rm o}$}

\newcommand{\Tripp}{3p~$\!^3$P$^{\rm o}$}

\begin{document}

\thesaurus{07 
    (02.01.4;  
     02.12.1;  
     08.01.1;  
     08.12.1)} 

\title{Non-LTE analysis of neutral magnesium in cool stars \thanks{Based on
observations collected at the German-Spanish Astronomical Center, Calar Alto,
Spain and Beijing Astronomical Observatory, Xinglong, China}}

\author{G. Zhao\inst{1,2} \and T. Gehren\inst{1,2}}

\offprints{G. Zhao}

\institute{%
Beijing Astronomical Observatory, Chinese Academy of Sciences, 100012 Beijing,
China \and Institut f\"ur Astronomie und Astrophysik der Universit\"at
M\"unchen, Scheinerstr. 1, 81679 M\"unchen, Germany\\ (zg@orion.bao.ac.cn,
gehren@usm.uni-muenchen.de)}
\date{Received 21 June 2000 / Accepted 31 August 2000}

\maketitle

\begin{abstract}

Calculations of the statistical equilibrium of magnesium in the solar
photosphere have shown that NLTE populations hardly affect Mg line formation in
the Sun. However, in metal-poor dwarfs and giants the influence of electron
collisions is reduced, and the ultraviolet radiation field, enhanced due to
reduced background line opacity, results in more pronounced NLTE effects. In the
photosphere of a cool star excitation and ionization due to collisions with
neutral hydrogen can outweigh electron collisions.  Analyses based on NLTE 
populations lead to significantly
higher Mg abundances than those calculated from LTE. We calculate magnesium
abundances in 10 cool dwarfs and subgiants with metallicities from $-2.29$ to
$0.0$. The results are based on spectra of high-resolution and high
signal-to-noise ratio. Stellar effective temperatures are derived from Balmer
line profiles, surface gravities from Hipparcos parallaxes and the wings of the
\ion{Mg}{i}b triplet, and metal abundances and microturbulence velocities are
obtained from LTE analyses of \ion{Fe}{ii} line profiles. For stars with
metallicities between $-2.0 <$ [Fe/H] $< -1.0$ abundance corrections $\Delta{\rm
[Mg/H]_{NLTE-LTE}} \sim 0.05 - 0.11$ are found. As expected the corrections
increase with decreasing metal abundance, and they increase slightly with
decreasing surface gravity. We also calculate the  statistical equilibrium of
magnesium for series of model atmospheres with different stellar parameters and
find that $\Delta{\rm [Mg/H]_{NLTE-LTE}}$ increases with effective temperature
between 5200 and 6500 K. For extremely metal-poor stars the abundance
corrections approach $\Delta{\rm [Mg/H]_{NLTE-LTE}} \sim 0.23$ at [Fe/H] $\sim
-3.0$.

\keywords{Atomic processes -- Lines: formation -- Stars: abundances --
 Stars: late-type}

\end{abstract}

\section{Introduction}

Abundance patterns of $\alpha$-elements, such as Mg, Si and Ca, in stars with
different metallicities are important for understanding the chemical evolution
of Galaxy. The first stellar generations are supposed to have produced mostly
$\alpha$-elements during massive supernova explosions of type II. Studies of
$\alpha$-element synthesis in SNe~II have been undertaken by Arnett
(\cite{ARNETT91}), Thielemann et al. (\cite{THIELEMANN96}), Woosley \& Weaver
(\cite{WOOSLEY95}), and Nakamura et al. (\cite{NAKA99}). Due to small
differences in the way stellar winds and semi-convection are treated and in the
specification of the {\em mass cut} and explosion mechanism their predictions
differ, mainly in the respective Fe yields but also in the pre-explosion yields
of Mg. There is no question that Mg, in principle, is less affected by the
fine-tuning of SN~II explosions. Therefore, inasmuch as the products of SN~II
nucleosynthesis are mixed into interstellar space \Mag\ should constitute a
reliable reference of the early evolution of the Galaxy. Due to the number of
strong absorption lines found in the spectra of even the metal-poor stars,
\ion{Mg}{i} is easier to observe than e.g. \ion{O}{i}. However, it shares the
disadvantage of most neutral metals in the atmospheres of moderately cool stars
with \ion{Mg}{ii} being the dominant ionization stage at temperatures above 5000
K. Consequently, neutral magnesium is sensitive to deviations from {\em local
thermodynamic equilibrium}, particularly as its ionization balance is dominated
by large photoionization cross-sections from the \Tripp\ state. In metal-poor
stars reduced line-blanketing produces an increased UV radiation field that, in
combination with the large cross-sections, leads to strong photoionization
rates; this makes \ion{Mg}{i} even more sensitive to NLTE and affects any
careful abundance analysis of Mg in these objects. Moreover, the density of free
electrons decreases roughly in proportion to metal abundance, and the collision
rates become correspondingly smaller. Together this change of microscopic
interaction processes should lead to substantially stronger deviations from LTE
in metal-poor stars at optical depths $\log$\,\tauc between $-3$ and $0$, if the
reduced electron collisional interaction is not compensated by hydrogen
collisions (Baum\"uller \& Gehren \cite{BAUM96},\cite{BAUM97}).

In our present study, we use several \ion{Mg}{i} lines chosen from the solar
spectrum which were carefully analyzed previously (Zhao et al. \cite{ZHAO98},
hereafter Paper~I). The standard atmospheric models we use in our present
analysis are described in Sect. 2. We start with test calculations including
the standard model atom of Paper I, and we fit the calculated NLTE equivalent
widths for models of different stellar parameters. We determine Mg {\em
abundance corrections} due to differences between NLTE and LTE analyses for
series of model atmospheres with the different stellar parameters in Sect. 3.
All \ion{Mg}{i} lines considered here are reproduced using standard NLTE line
formation techniques with the radiative transfer solved in the Auer-Heasley
scheme, taking into account the interaction processes between all levels of the
Mg model atom. A previous investigation (Paper~I) has demonstrated how the line
profiles observed in the solar spectrum can be used to improve the atomic model
of neutral magnesium. This model is the subject of further scrutiny in the light
of information coming from comparison of test calculations with line profiles
in the spectra of metal-poor stars. We then calculate magnesium abundances of 
10 cool dwarfs and subgiants with metallicities from $-2.29$ to $0.0$ using both
LTE and NLTE line formation techniques in Sect. 4. Our conclusions are given
in last section.
\begin{figure}
\resizebox{\hsize}{!}{\includegraphics{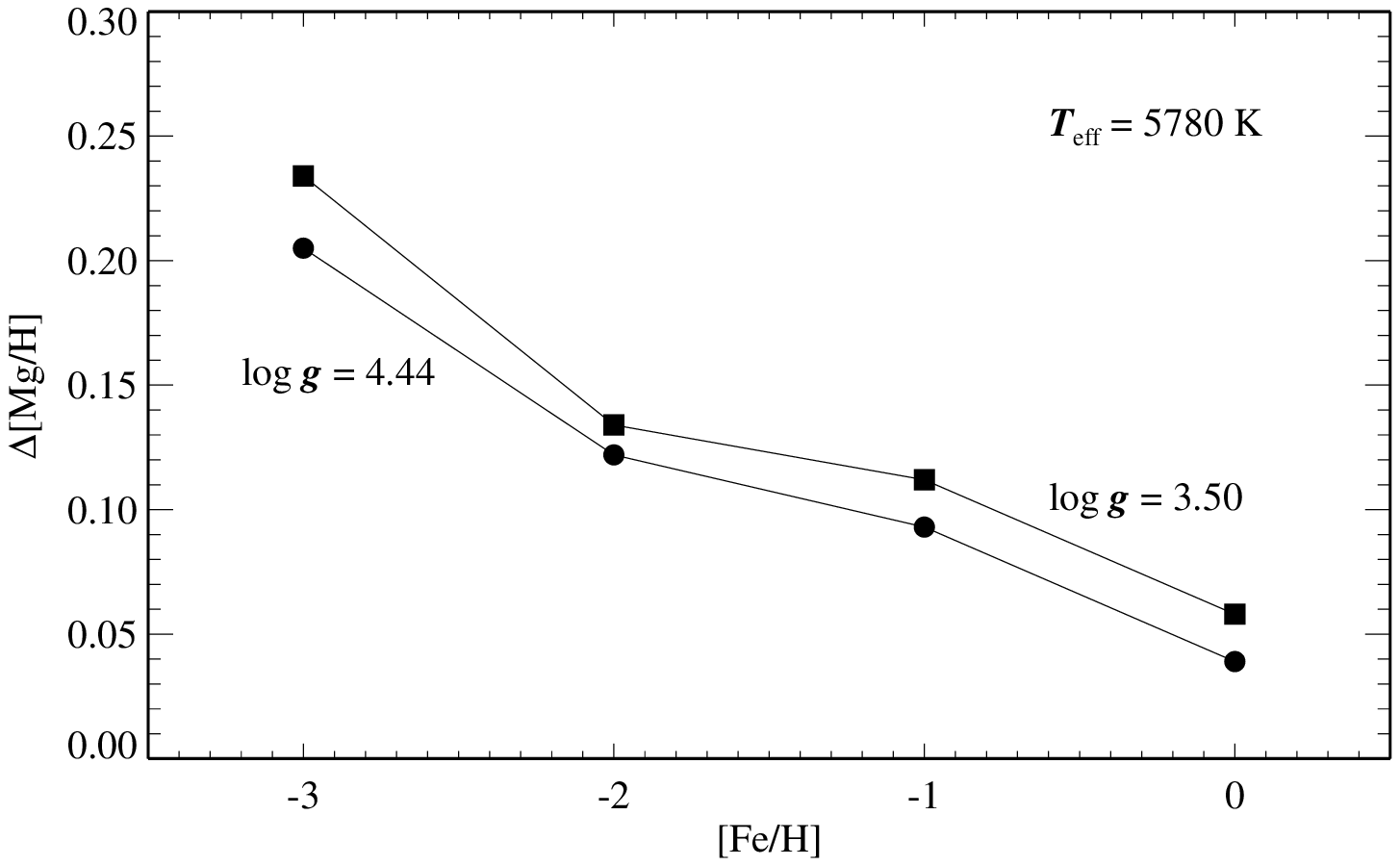}}\\ 
\resizebox{\hsize}{!}{\includegraphics{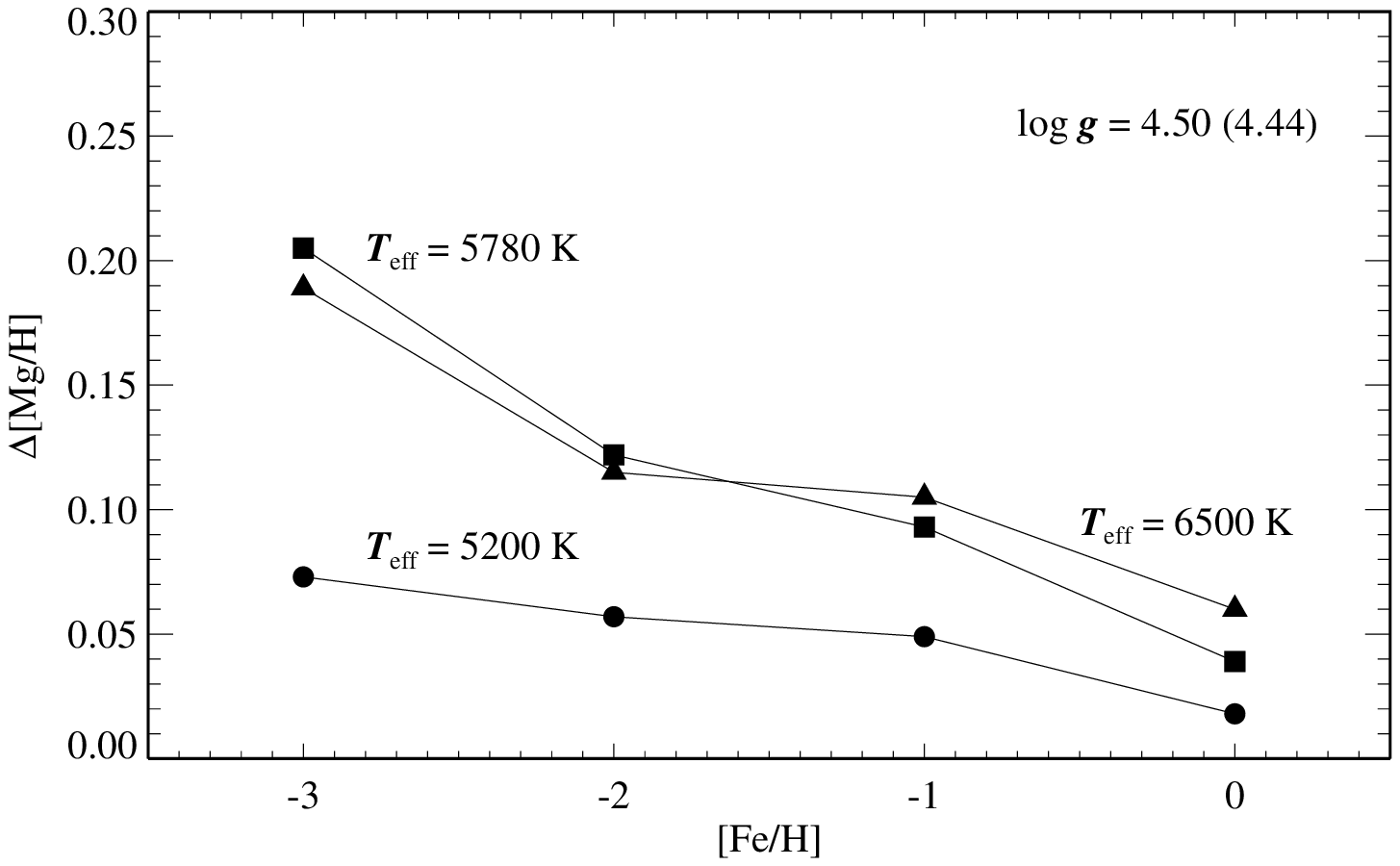}}   
\caption[]{NLTE corrections of the magnesium abundances versus [Fe/H] for the
model grids, where $\Delta$[Mg/H] = [Mg/H]$_{\rm NLTE}$ -- [Mg/H]$_{\rm LTE}$.
Top: \Teff = 5780 K, \logg = 3.50 and 4.44. Bottom: \logg = 4.50 (4.44 for \Teff
= 5780 K), \Teff = 5200, 5780, and 6500 K.}
\label{nlte_corr}
\end{figure}
\begin{table*}
\caption[]{Abundance differences between NLTE and LTE obtained by fitting
calculated LTE equivalent widths of \ion{Mg}{i} lines using the same parameters
for various stellar model atmospheres based on our line-blanketed model grid. 
Results refer to $\Delta$[Mg/H] = 
[Mg/H]$_{\rm NLTE}$ - [Mg/H]$_{\rm LTE}$} \label{nltediff}
\begin{tabular}{lrrrrrrrrrrrr}
\hline \noalign{\smallskip}
\Teff& \logg &[Fe/H]&4571\AA&4703\AA&4730\AA&5528\AA&5711\AA&7657\AA&8806\AA&8213\AA&11828\AA&$\overline{\Delta[Mg/H]}\pm\sigma$\\
\noalign{\smallskip} \hline \noalign{\smallskip}
5200 &  4.50 &  0.0 & 0.022 & 0.014 & 0.021 & 0.010 & 0.015 & 0.035 & 0.001 & 0.042 & 0.004  & 0.018$\pm$0.014\\
5200 &  4.50 & -1.0 & 0.060 & 0.038 & 0.051 & 0.022 & 0.045 & 0.065 & 0.019 & 0.115 & 0.025  & 0.049$\pm$0.030\\
5200 &  4.50 & -2.0 & 0.091 & 0.039 & 0.058 & 0.028 & 0.052 & 0.055 & 0.045 & 0.098 & 0.045  & 0.057$\pm$0.023\\
5200 &  4.50 & -3.0 & 0.113 & 0.062 & 0.068 & 0.058 & 0.060 & 0.050 & 0.075 & 0.095 & 0.075  & 0.073$\pm$0.020\\[2mm]

5780 &  4.44 &  0.0 & 0.047 & 0.028 & 0.042 & 0.023 & 0.050 & 0.060 & 0.004 & 0.076 & 0.019  & 0.039$\pm$0.022\\
5780 &  4.44 & -1.0 & 0.087 & 0.073 & 0.088 & 0.060 & 0.104 & 0.130 & 0.041 & 0.202 & 0.050  & 0.093$\pm$0.049\\
5780 &  4.44 & -2.0 & 0.128 & 0.123 & 0.111 & 0.099 & 0.117 & 0.120 & 0.079 & 0.234 & 0.090  & 0.122$\pm$0.045\\
5780 &  4.44 & -3.0 & 0.182 & 0.190 & 0.172 & 0.180 & 0.172 & 0.200 & 0.163 & 0.430 & 0.157  & 0.205$\pm$0.085\\[2mm]

5780 &  3.50 &  0.0 & 0.060 & 0.033 & 0.062 & 0.028 & 0.090 & 0.115 & 0.016 & 0.095 & 0.026  & 0.058$\pm$0.035\\
5780 &  3.50 & -1.0 & 0.101 & 0.081 & 0.120 & 0.054 & 0.151 & 0.180 & 0.018 & 0.252 & 0.053  & 0.112$\pm$0.073\\
5780 &  3.50 & -2.0 & 0.147 & 0.157 & 0.137 & 0.118 & 0.136 & 0.120 & 0.045 & 0.264 & 0.085  & 0.134$\pm$0.060\\
5780 &  3.50 & -3.0 & 0.201 & 0.224 & 0.206 & 0.211 & 0.207 & 0.210 & 0.212 & 0.429 & 0.210  & 0.234$\pm$0.073\\[2mm]

6500 &  4.50 &  0.0 & 0.042 & 0.041 & 0.046 & 0.045 & 0.083 & 0.100 & 0.042 & 0.095 & 0.047  & 0.060$\pm$0.025\\
6500 &  4.50 & -1.0 & 0.077 & 0.102 & 0.088 & 0.090 & 0.123 & 0.150 & 0.051 & 0.188 & 0.072  & 0.105$\pm$0.043\\
6500 &  4.50 & -2.0 & 0.096 & 0.142 & 0.101 & 0.120 & 0.105 & 0.120 & 0.075 & 0.178 & 0.094  & 0.115$\pm$0.031\\
6500 &  4.50 & -3.0 & 0.148 & 0.178 & 0.162 & 0.182 & 0.179 & 0.200 & 0.178 & 0.302 & 0.175  & 0.189$\pm$0.045\\
\noalign{\smallskip}
\hline
\end{tabular}
\end{table*}

\section{Model atmospheres}

The statistical equilibrium calculations are performed in horizontally
homogeneous LTE model atmospheres in hydrostatic equilibrium. We account for
metallic and molecular UV line absorption using Kurucz' (\cite{KURUCZ92})
opacity distribution functions (ODF) interpolated for the proper stellar
metallicities. In order to use the meteoritic iron abundance $\log
\varepsilon_{Fe} = 7.51$ as the solar reference, we scaled Kurucz' ODF by
$-0.16$ dex to put his iron opacity on the proper scale. Bound-free opacities
were computed with solar abundances from Holweger (\cite{HOLWEGER79}), scaled by
the stellar metallicities, and in case of [Fe/H] $< -0.6$ corrected for a
general enhancement of $\alpha$-element abundances of O, Mg, and Si with
[$\alpha$/Fe] = 0.4. Since Mg is the most important electron donator among
the $\alpha$-elements this approximation follows the results found by
Fuhrmann et al. (1995). Convection is taken into account parametrically 
with a mixing-length of 0.5 pressure scale heights (see Fuhrmann et al. 
\cite{FAG93}).

The reason for adopting this model as a standard is that we can easily use the
model to calculate stellar atmospheres differentially with full physics
included. On the other hand, the abundance {\em differences} between the LTE and
NLTE calculations have nearly the same value independent of the type of model
atmospheres adopted, even though we had to adjust some line parameters (e.g.
logC$_{6}$, \loggf) in order to obtain a better line fit (\cf Paper~I).

\section{NLTE line formation}

\subsection{The standard atomic model}

The standard atomic model we used in this analysis is basically the same as that
used for the solar calculations (\cf Paper I). The neutral magnesium model
includes all levels n$\ell$ up to n $= 9$ and $\ell \le \rm{n} - 1$ resulting in
a total of 83 \ion{Mg}{i} terms plus the doublet ground state of \ion{Mg}{ii}.
All energies were taken from the compilation of Martin \& Zalubas
(\cite{MART80}) except for those terms with $\ell>5$ which were obtained using
the polarization formula of Chang \& Noyes (\cite{CHANG83}). Fine structure
splitting has been neglected. This model is nearly the same as that used by
Carlsson et al. (\cite{CARL92}). All terms are coupled by radiative and
collisional interactions as described in more detail in Paper I.

We have shown in Paper~I that collisional interaction with neutral hydrogen
atoms will not affect the results obtained for the solar \ion{Mg}{i} lines
except those in the infrared. However, since neutral hydrogen collision rates do
not depend on metal abundance, they can maintain an important influence in
metal-poor stellar atmospheres and raise the total collision rates to high
enough efficiency at which they can even compensate the strongly increased
photoionization. As outlined in Paper~I the hydrogen collisions are based on the
formula of Drawin (\cite{DRAWIN69}). It is most often used in a form due to
Steenbock \& Holweger (\cite{STEE84}) although the validity of this formula
cannot be judged. We have used it as an order of magnitude estimate and
carefully investigated the influence of hydrogen collisions on our atomic model
by fitting all available \ion{Mg}{i} lines modifying Drawin's formula by a
scaling factor. For \ion{Mg}{i} it turns out (as was the case for aluminium)
that this factor should vary {\em exponentially} with upper level excitation
energy $E_{\rm n}$ (in eV), $S_{\rm H} = 1000 \exp\{-{\rm n}E_{\rm n}/2\}$. This
was determined in a fully empirical manner recomputing the complete NLTE line
formation with statistical equilibrium equations including the scaled hydrogen
collision rates, and it enabled us to fit solar lines of different excitation
energies. The decrease of hydrogen collision cross-sections with excitation
energy is also in rough agreement with Kaulakys'
(\cite{KAULAKYS85}, \cite{KAULAKYS86}) prediction for Rydberg transitions.

\subsection{Statistical equilibrium calculations for different stellar parameters}

The statistical equilibrium is calculated using the {\sc Detail} code (Giddings
\cite{GIDDINGS81}; Butler \& Giddings \cite{BUTLER85}) in a version based on the
method of complete linearization as described by Auer \& Heasley
(\cite{AUER76}). The calculation includes all radiative line transitions, most
of them represented by Doppler profiles; 99 lines were linearized. The Mg b
lines are treated with full radiative and van der Waals damping. The linearized
line transitions were selected from test calculations including different
combinations with a preference for the stronger transitions including the $\ell
= \rm{n} - 1$ levels. Adding more transitions did not change the results. The
bound-free transitions of the lowest 22 levels were linearized, too.

In a star such as the Sun the flux in the ultraviolet spectral region is
determined to a large part by opacities due to metal line absorption. As in our
analysis of the solar Mg spectrum we use the opacity distribution functions of
Kurucz (\cite{KURUCZ92}) to represent this opacity. In these ODF data single
line opacities in small frequency intervals are represented by superlines;
consequently, the ODF opacity is not identical to that required at a specific
position in frequency space. For the calculation of a model atmosphere this
simplification is a sufficient approximation, but for NLTE line formation the
actual radiation field across a line transition or an ionization continuum is
important for the determination of the statistical equilibrium of an atom. For
bound-free transitions the exact position of the absorbing lines is less
important, and the use of the ODFs will be reliable, provided the intervals are
small enough in the frequency region near the ionization edge. For a bound-bound
transition with its narrow line width it can be important in which part of the
broad synthetic ODF line it is formed. We include the additional ODF opacity in
the UV for wavelengths between 1300 and 3860 \AA~ to allow for a realistic
behaviour of the ionization from the ground state and the first excited level
without affecting most of the line transitions. Only a few \ion{Mg}{i} lines are
found in this region allowing us to omit additional line opacities as most of
these transitions are of minor importance for the statistical equilibrium.

We use the statistical equilibrium of \ion{Mg}{i} with hydrogen collisions from
Drawin's formula (Drawin \cite{DRAWIN69}) but scaled exponentially with upper
level excitation energy as described in Sect. 3.1. Abundance differences
between the NLTE and LTE calculations with various stellar models are shown in
Table \ref{nltediff}. These results are obtained requiring NLTE and LTE line
formation to fit the same equivalent width. The results confirm that deviations
of the \ion{Mg}{i} level populations from LTE increase with decreasing metal
abundance. The NLTE effects in solar metallicity stars are almost negligible. In
Fig. \ref{nlte_corr} we demonstrate the variation of the NLTE corrections of
magnesium abundances versus [Fe/H] for sequences of model grids with constant
surface gravity, \logg $\sim 4.5$, and \Teff $= 5200, 5780$ and $6500$ K, and with
constant effective temperature, \Teff $= 5780$ K, and \logg $= 3.5$ and $4.44$.
From these plots it is evident that the NLTE effects are systematically stronger
for the hotter models, which is in accordance with the statistical equilibrium
of aluminium and sodium (Baum\"uller \& Gehren \cite{BAUM96}; Baum\"uller et al.
\cite{BAUM98}). The dependence upon temperature is, however, strongly tied to
the change in ionization; as soon as \Teff drops below 5300 K, \ion{Mg}{ii}
starts to recombine, and \ion{Mg}{i} becomes more dominant. For stars at the
turnoff the temperature dependence is negligible. As expected, the strongest
departures from LTE are found for models with high temperature and low
metallicity. The reduction of surface gravity results in a decreased efficiency
of collisions by electron and hydrogen atoms which again leads to slightly
stronger NLTE effects.
\begin{figure}
\resizebox{\hsize}{!}{\includegraphics{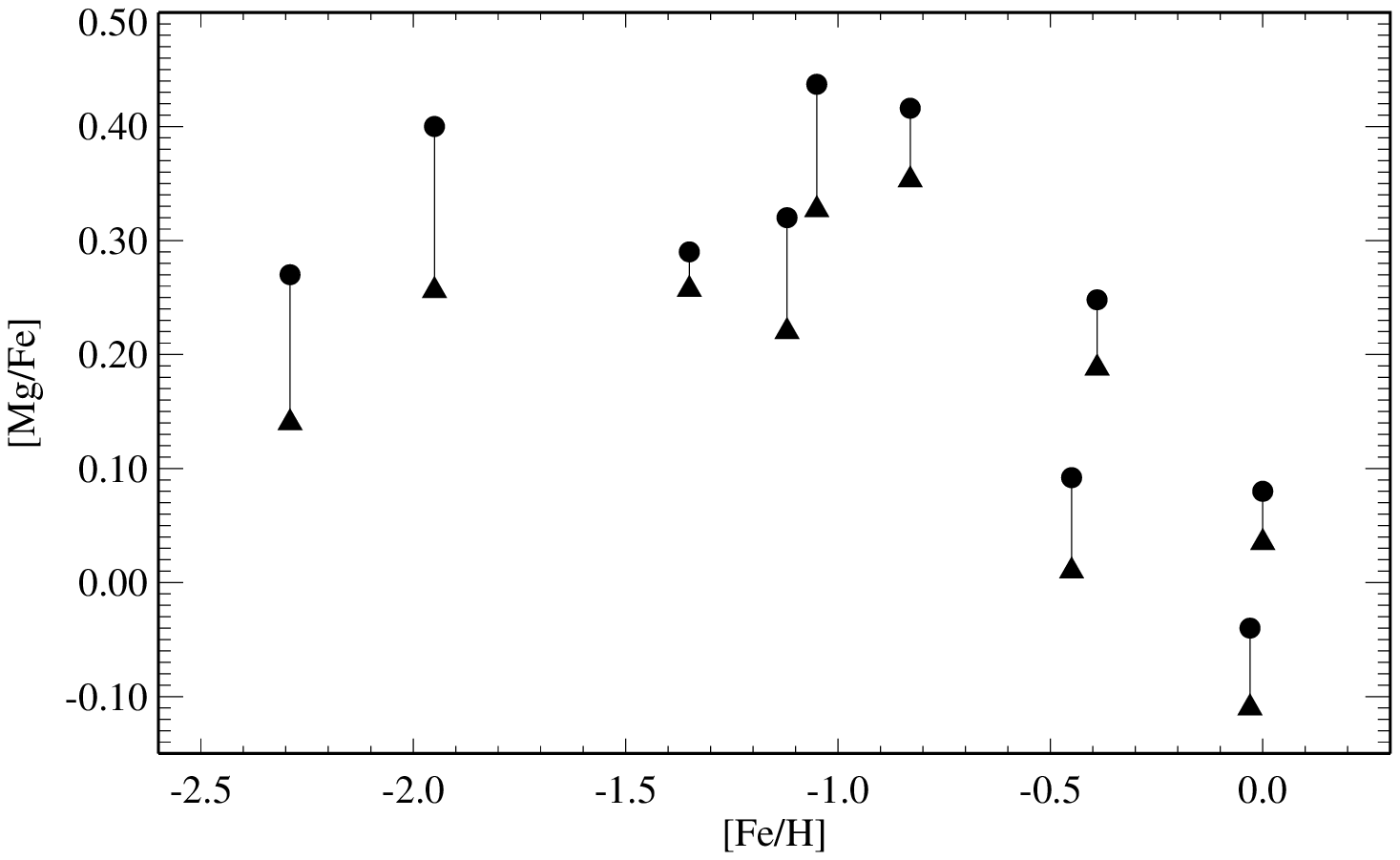}}\\ 
\resizebox{\hsize}{!}{\includegraphics{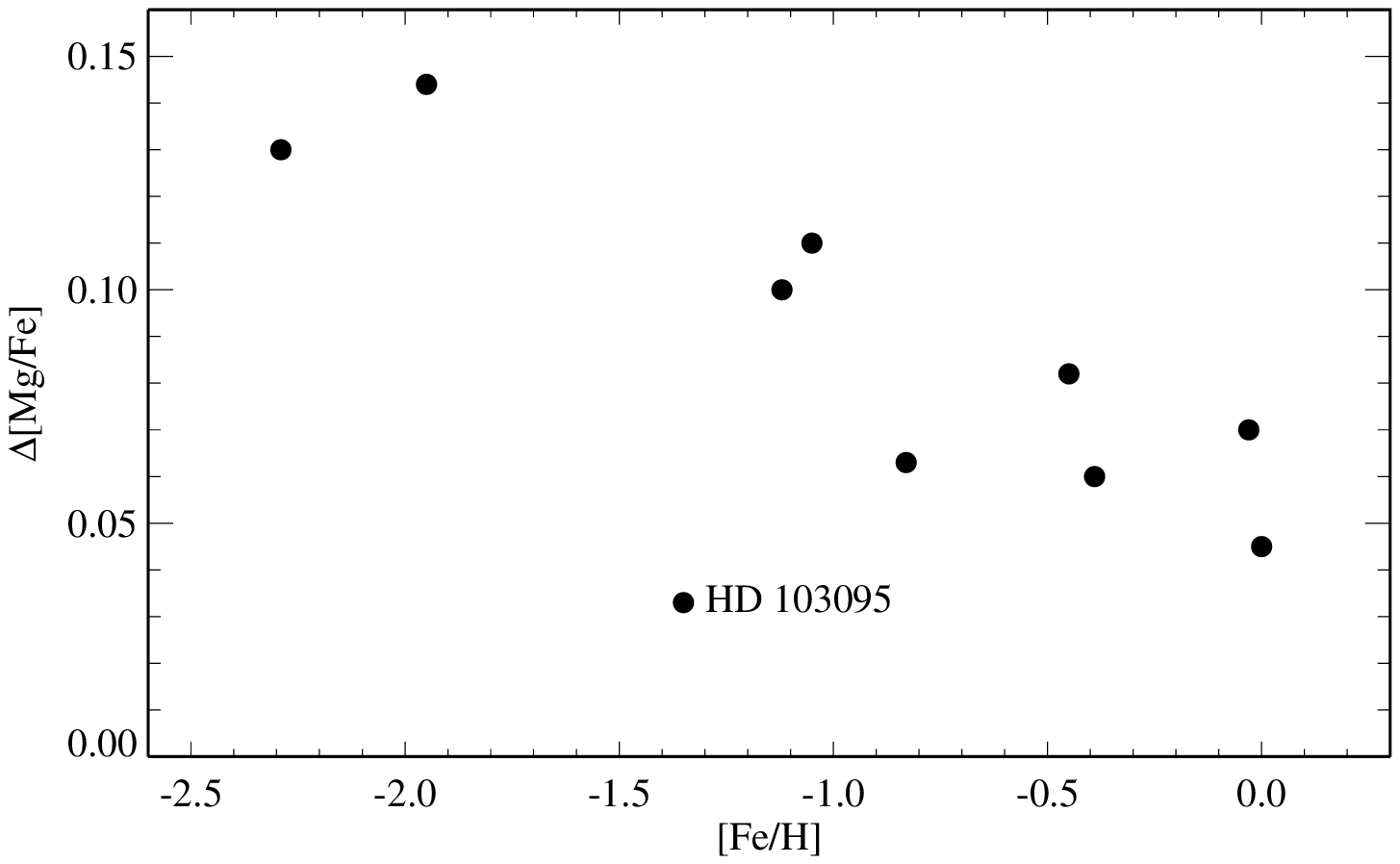}}\\ 
\resizebox{\hsize}{!}{\includegraphics{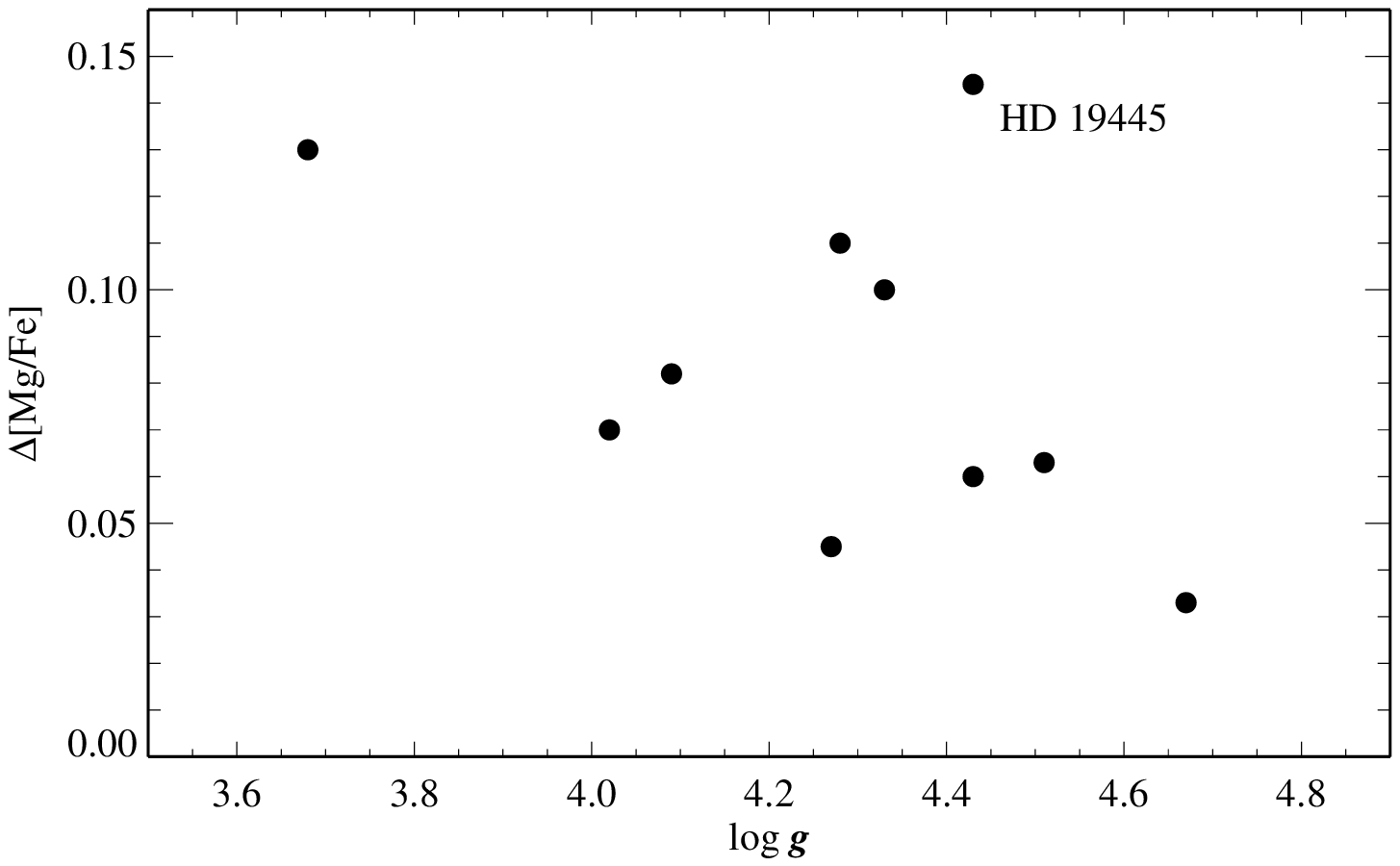}}   
\caption[]{Top: Magnesium abundances of 10 cool dwarfs and subgiants from NLTE
and LTE line formation. The symbols of triangle and filled circle represent the
LTE and NLTE results, respectively. NLTE corrections $\Delta$[Mg/Fe] of the 10
program stars versus [Fe/H], where $\Delta$[Mg/Fe] = [Mg/Fe]$_{\rm NLTE}$ -
[Mg/Fe]$_{\rm LTE}$ as a function of metal abundance (middle) and surface gravity
(bottom)}
\label{mgabund}
\end{figure}

\section{Magnesium abundances in cool stars}

Our present study is based on spectroscopic data obtained by K. Fuhrmann and M.
Pfeiffer using the fiber optics cassegrain echelle spectrograph (FOCES) fed by
the 2.2m telescope at the Calar Alto Observatory during three observing runs in
September 1995, October 1996 and May 1997. The 1995 spectra were exposed to a
1024 $\times$ 1024 24$\mu$ CCD, and the resolving power was $\sim$ 35,000. In
October 1996 and May 1997 a 2048 $\times$ 2048 15$\mu$ CCD was employed at R
$\sim$ 60,000. Almost all the stars were observed at least twice with the
signal-to-noise ratio of about 200 (Fuhrmann \cite{F98}). Details with respect
to the FOCES spectrograph can be found in Pfeiffer et al. (\cite{PFE98}). For
HD\,19445, HD\,95128, HD\,103095, HD\,194598, and HD\,201891 additional spectra
were obtained by G. Zhao and H. Zhang with the Coud\'{e} Echelle Spectrograph
attached to the 2.16 meter telescope at Beijing Astronomical Observatory
(Xinglong, China) in October 1996 and March 1997. The detector is a Tek CCD
(1024$\times$1024 pixels of 24 $\mu$m$^{2}$ each). The Coud\'{e} Echelle
Spectrograph has two beams: a blue path and a red path. For the blue path, the
79 grooves/mm echelle grating is used along with the prism as cross-disperser
and a 0.5 mm slit, leading to a resolving power of the order of 44,000. For the
red path, the 31.6 grooves/mm echelle grating is used along with the prism as
cross-disperser and a 0.5 mm slit, leading to a resolving power of the order of
37,000. The more detailed technique and performance descriptions of the
Coud\'{e} Echelle Spectrograph can be found in the paper of Jiang
(\cite{JIANG96}).

\begin{table*}
\caption[]{Stellar parameters and magnesium abundances [Mg/Fe] calculated with
LTE and NLTE line formation}
\begin{tabular}{rrrrrrrrrrrrrr}
\hline
\noalign{\smallskip}
Object &   V~~ & \Teff & \logg & ~[Fe/H] & $\xi_{t}$~~ &
\multicolumn{2}{c}{4571 \AA} & \multicolumn{2}{c}{4703 \AA} &
\multicolumn{2}{c}{5528 \AA} & \multicolumn{2}{c}{5711 \AA} \\
  HD~~ & [mag] &  [K]~ & & & ~[km/s] & LTE & NLTE & LTE & NLTE & LTE & NLTE & LTE & NLTE \\
\noalign{\smallskip} \hline \noalign{\smallskip}
  6582 &  5.16 & 5387  & 4.51 &  --0.83 &  0.89  &      - &      - & 0.33 & 0.39 &   0.36 & 0.43 &   0.37 &   0.43 \\
 19445 &  8.06 & 6016  & 4.43 &  --1.95 &  1.35  &   0.38 &   0.48 & 0.13 & 0.36 &   0.26 & 0.36 &      - &      - \\
 30743 &  6.26 & 6298  & 4.09 &  --0.45 &  1.64  & --0.02 &   0.04 &    - &    - &   0.07 & 0.14 & --0.02 &   0.10 \\
 61421 &  0.37 & 6470  & 4.02 &  --0.03 &  1.95  & --0.19 & --0.14 &    - &    - & --0.01 & 0.03 & --0.13 & --0.01 \\
 95128 &  5.05 & 5892  & 4.27 &  --0.00 &  1.01  &      - &      - &    - &    - &   0.02 & 0.04 &   0.05 &   0.12 \\
103095 &  6.45 & 5110  & 4.67 &  --1.35 &  0.85  &   0.21 &   0.27 & 0.26 & 0.28 &   0.31 & 0.32 &   0.25 &   0.29 \\
140283 &  7.21 & 5810  & 3.68 &  --2.29 &  1.49  &      - &      - &    - &    - &   0.14 & 0.27 &      - &      - \\
165401 &  6.81 & 5811  & 4.43 &  --0.39 &  1.10  &   0.17 &   0.23 & 0.17 & 0.23 &   0.23 & 0.27 &   0.18 &   0.26 \\
194598 &  8.34 & 6058  & 4.33 &  --1.12 &  1.45  &   0.25 &   0.33 & 0.20 & 0.32 &   0.24 & 0.32 &   0.19 &   0.31 \\
201891 &  7.37 & 5943  & 4.28 &  --1.05 &  1.18  &      - &      - & 0.31 & 0.43 &   0.38 & 0.45 &   0.29 &   0.43 \\
\noalign{\smallskip} \hline
\end{tabular}
\label{tabpar}
\end{table*}

The effective temperatures of our program stars were determined from the wings
of the Balmer lines H$_{\alpha}$ and H$_{\beta}$ profile fitting. For metal-poor
stars, H$_{\gamma}$ and H$_{\delta}$ are also taken into account (\cf Fuhrmann
\cite{F98}). Fortunately, the dependence of the results on effective temperature 
is not important for most of the stars; only HD\,6582 and HD\,103095 are cool 
enough to be affected (see Fig. \ref{nlte_corr}). Surface gravities have been 
derived from Hipparcos data using the method given by Nissen \& Schuster 
(\cite{NISS97}). The final surface gravities are determined by comparison to the 
strong line wings of the \ion{Mg}{i}b triplet which gives a slightly higher 
value than that from the determination of Hipparcos data. The [Fe/H] values were 
adopted from the determination of \ion{Fe}{ii} lines which are insensitive to 
NLTE effects (Th\'evenin \& Idiart \cite{TI99}). The mircoturbulence velocity 
$\xi_{t}$ was estimated using an extended set of Fe lines. 

\begin{figure*}
\resizebox{\textwidth}{!}{\includegraphics{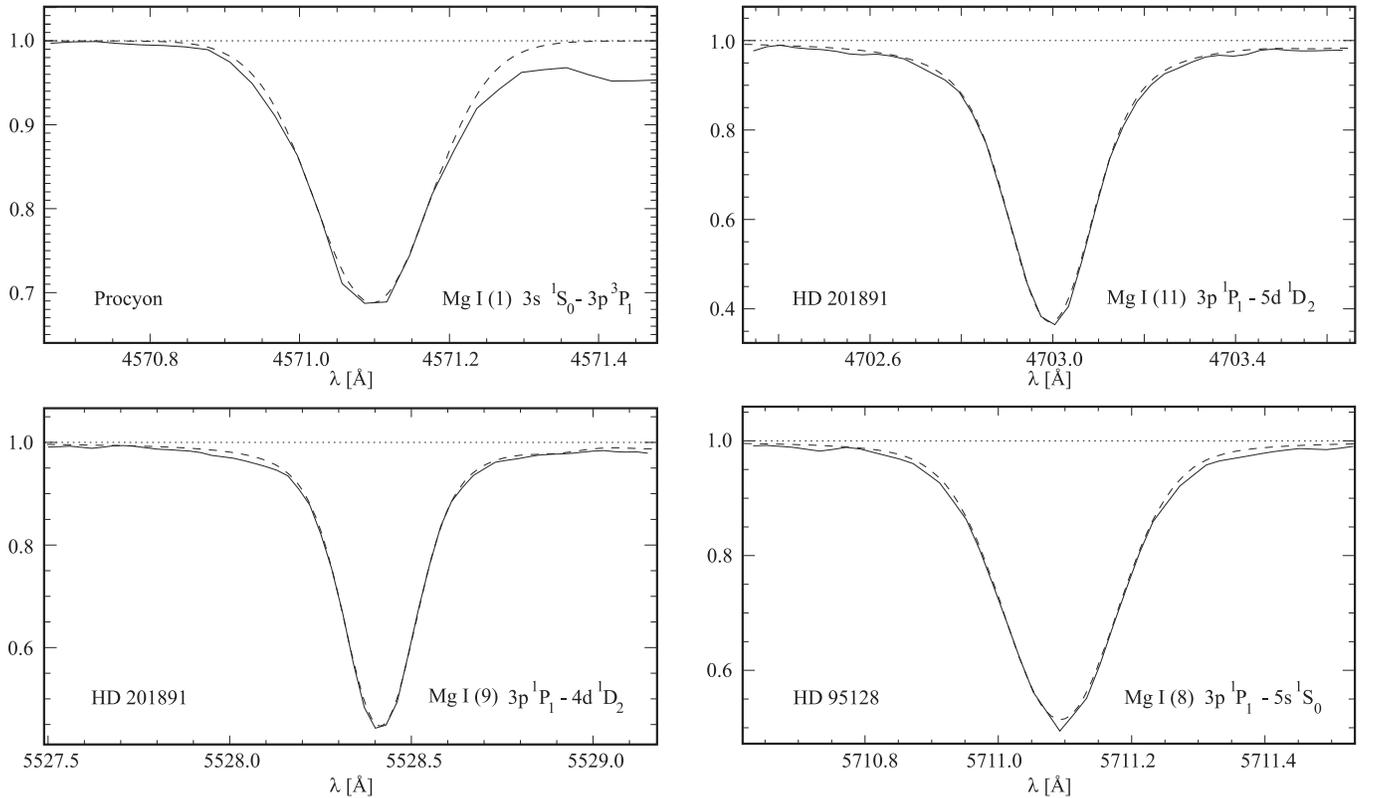}} \caption[]{Synthetic 
flux profiles of selected \ion{Mg}{i} lines (dashed) compared with the observed 
spectra of program stars (continuous). The synthetic line profiles refer to NLTE 
with hydrogen collisions scaled exponentially with excitation energy. ({\em top 
left}): intercombination line \Sings~-- \Tripp~ at 4571 \AA . ({\em top right}): 
excited line \Singp~ -- 5d~$\!^1$D transition at 4703 \AA . ({\em bottom left}): 
excited line \Singp~ -- 4d~$\!^1$D transition at 5528 \AA . ({\em bottom 
right}): excited line \Singp~ -- 5s~$\!^1$S transition at 5711 \AA .} 
\label{sing_s} \label{profiles} 
\end{figure*}

We note that only a few \ion{Mg}{i} lines show a suitable line profile for
spectrum synthesis in our spectra of sample stars. Therefore, four lines, namely
$\lambda\lambda$ 4571, 4703, 5528 and 5711\AA, which are often used for
magnesium abundance determinations are selected in our present analysis.
Synthetic spectra using either LTE or NLTE level populations were calculated
using the line data given in Paper~I. The final results are presented in Table
\ref{tabpar}, where both the NLTE and the LTE line profile fit abundances are
listed for comparison. We present the magnesium abundances of 10 cool dwarfs and
subgiants using NLTE and LTE line formation in Fig. \ref{mgabund}; the plots of
the NLTE correction $\Delta$[Mg/Fe] versus [Fe/H] and \logg are shown, too. One
of the most striking results is the difference between NLTE and LTE abundances
in the very metal-poor subgiant HD\,140283 and subdwarf HD\,19445, corresponding
to factors of 1.35 and 1.39, respectively. It is again evident how NLTE
abundance corrections increase with decreasing stellar metallicity. There is one
star HD\,103095 (Groombridge 1830) that shows a different behaviour in Fig. 
\ref{mgabund} (middle); due to its low effective temperature ($\sim 5110$ K, the 
lowest one in our sample stars) its ionization equilibrium is more affected by 
temperature than by NLTE effects. This is in accordance with our analysis for 
the grids of the model atmospheres in Sect. 3. The effective temperature, 
surface gravity, metallicity and microturbulence have typical error of 
$\Delta$\Teff $\sim$ 80~K, $\Delta$\logg $\sim$ 0.1 dex, $\Delta$[Fe/H] $\sim$
0.07 dex and $\Delta\xi_{t}$ $\sim$ 0.2 km/s. Consequently, the error estimate
for the magnesium over hydrogen abundances is about 0.07 dex. These error 
estimations are nearly the smae for LTE and NLTE abundance determination.
Note that in most cases the abundance differences are well above the 
observational errors. On the other 
hand, the NLTE abundances based on various line profile fits generally produce a 
satisfactory smaller standard deviation than that of LTE abundances. This result 
is perhaps the most important because it shows that with high resolution ($R > 
40000$), high signal-to-noise spectra it is possible to reproduce line profiles 
with very high accuracy. A few profile fits for the different magnesium lines 
are given in Fig. \ref{profiles} to show the quality of the line synthesis. 
They are representative for the average fit quality obtained for the $R \sim 
40000$ and $60000$ spectra.\\

\section{Conclusions}

The variation of [Mg/Fe] with the stellar mean metallicity [Fe/H] contains 
information about the chemical evolution of the Milky Way. Previous 
investigations of magnesium based on LTE analyses by a number of researchers 
(\eg Magain \cite{MAGAIN87}; Hartmann \& Gehren \cite{HARTMANN88}; Fuhrmann et 
al. \cite{FUHRMANN95}; Fuhrmann \cite{F98}) show that magnesium has a solar 
abundance ratio in disk stars ([Fe/H] $\geq$ $-$0.5), and then [Mg/Fe] becomes 
constant at +0.3 \ldots 0.4 in metal-poor stars ([Fe/H] $\leq  -0.5$). As shown 
by Fuhrmann (\cite{F98}), this overabundance is a distinct result of population 
membership. Our new results based on the NLTE analysis of 10 cool unevolved 
stars demonstrate that the magnesium overabundance with respect to iron becomes 
even larger by roughly 0.1 dex for stars with reduced stellar metal abundance 
[Fe/H] due to photoionization dominating over collisions. The reduction of the 
surface gravity results in a decreased efficiency of collisions by electron and 
hydrogen atoms which also leads to small NLTE effects. The strongest departures 
from LTE are found for metal-poor stars with high temperature and low gravity, 
i.e. {\em turnoff} stars. 

Fortunately, the trend obtained from LTE abundance analyses of magnesium in cool 
stars (Fuhrmann \cite{F98}) is confirmed. The significant [Mg/Fe] ratio in 
metal-poor stars is found from NLTE spectrum analyses, with the increased 
level indicating a shift to even higher stellar masses in the parent population 
(III). Such a trend was also found by Fuhrmann et al. (1995) or Audouze \& Silk 
(1995) who investigated the initial mass function based on the
$\alpha$-element enrichment of population II.
Recent investigations of the \ion{Fe}{ii}/\ion{Fe}{i} equilibrium (see 
Th\'evenin \& Idiart \cite{TI99}), however, seem to demonstrate that [Mg/Fe] may 
be more affected by deviations from LTE when using \ion{Fe}{i} linese
to derive iron abundances. We recall that in our analysis we used only 
\ion{Fe}{ii} linese which are insensitive to NLTE effects. 

In order to confirm the present results for magnesium, and in particular the 
mean abundance differences between LTE and NLTE it is necessary to reinvestigate 
a larger sample of stars with various metallicities based on the NLTE analysis, 
especially for the metal-poor stars of the thick disk and halo. Such an 
investigation should, however, include an \ion{Fe}{i} NLTE analysis. Then it 
will be an important step improving our present understanding of stellar 
nucleosynthesis and the chemical evolution of the Galaxy.

\acknowledgements{This research was supported by the Deutsche
Forschungsgemeinschaft, the National Natural Science Foundation 
of China, and the Major State Basic Research Development Program of China. 
We thank Dr. Klaus Fuhrmann who kindly provided the spectra data of the
program stars. We thank Dr. Johannes Reetz for supporting the line 
synthesis code.}

\end{document}